# Fixed Inducing Points Online Bayesian Calibration for Computer Models with an Application to a Scale-Resolving CFD Simulation

Yu Duan[1*], Matthew Eaton[1], Michael Bluck[1]

1. Nuclear Engineering Group/Rolls-Royce Modelling and Analytics (M&A) Group, Department of Mechanical Engineering, Imperial College London, SW7 2BX, UK.

Abstract

This paper proposes a novel fixed inducing points online Bayesian calibration (FIPO-BC) algorithm to efficiently learn the model parameters using a benchmark database. The standard Bayesian calibration (STD-BC) algorithm provides a statistical method to calibrate the parameters of computationally expensive models. However, the STD-BC algorithm scales very badly with the number of data points and lacks online learning capability. The proposed FIPO-BC algorithm greatly improves the computational efficiency and enables the online calibration by executing the calibration on a set of predefined inducing points.

To demonstrate the procedure of the FIPO-BC algorithm, two tests are performed, finding the optimal value and exploring the posterior distribution of 1) the parameter in a simple function, and 2) the high-wave number damping factor in a scale-resolving turbulence model (SAS-SST). The results (such as the calibrated model parameter and its posterior distribution) of FIPO-BC with different inducing points are compared to those of STD-BC. It is found that FIPO-BC and STD-BC can provide very similar results, once the predefined set of inducing point in FIPO-BC is sufficiently fine. But, the FIPO-BC algorithm is at least ten times faster than the STD-BC algorithm. Meanwhile, the online feature of the FIPO-BC allows continuous updating of the calibration outputs and potentially reduces the workload on generating the database.

---

[1*] Corresponding author. Email: y.duan@imperial.ac.uk

# 1 Introduction

Computational modelling is now an integral tool to predict and understand complex physical systems in many branches of science and engineering. Computational modelling enables engineers and scientists to analyse physical phenomena which cannot be easily measured by experiments and reduces the time and financial cost of analysing complex systems. However, a computational model is normally subject to various types of errors, such as discretisation error, parametric uncertainty error, and model error. The discretisation error, parametric uncertainty error and model error all contribute to the total error in a mathematical or computational model. It is only by assessing and reducing each of the components to the total error that one can assess the accuracy of a given mathematical or computational model. The discretisation and parametric uncertainty errors are relatively easy to reduce, depending upon the computational resources and experimental data available. However, the model error is normally linked to the underlying mathematical model not representing the exact physics of the system being analysed. Therefore, model error is particularly challenging to both rigorously analyse and reduce in magnitude. In many applications, such as turbulence modelling, the model parameter is determined based upon approximations or simplified conditions. Therefore, model parameter calibration with respect to (w.r.t.) the benchmark data becomes important to improve the performance of the model.

The determination of the model parameters through calibration, data adjustment or data assimilation enables the model more accurately represent the underlying physics of a system being analysed. By this definition, it is natural to calibrate the model parameters using the best-fit principle. However, the best-fit principle does not consider the uncertainties that the data may be subjected to, such as the observational/numerical discretisation errors, which should be considered in determining the agreement between two databases [1–4]. Kennedy and O'Hagan [5] proposed a Bayesian approach to calibrating the computational model by representing model bias and computer model outputs as the Gaussian processes. The method automatically integrates the uncertainties into the calibration by treating them as Gaussian noises. For simplicity, we will refer to Kennedy and O'Hagan's approach as standard Bayesian Calibration (STD-BC).

Since the development of STD-BC, the concept was adopted and further developed by many researchers in different areas of science and engineering. For example, Bayarri et al [6–8] applied the method to validate models for resistance spot welding, dynamic stress and vehicle collision modelling. In computational fluid dynamic (CFD) simulation, it was applied to tuning the model parameters and inferring the true physical value [9,10], quantifying the inverse uncertainty [11–18] and informing the design of supercomputer simulations [19]. Recently, Wu et al. [20,21] modularised the STD-BC by dividing the database into two different sets; a calibration set and an inverse uncertainty quantification set. This framework has been successfully applied in Liu et al [22] to carry out the inverse uncertainty quantification of the correlation in the two-fluid model based multiphase-CFD simulation. Wu et al. [23] furtherly discussed the "lack of identifiability" issue of inverse uncertainty quantification using the Bayesian calibration. Karagiannis and Lin [24] proposed a mixture Bayesian Calibration framework, which can represent the real system by mixture the outputs of available computer models with different fidelity level.

Despite the successful applications listed above, the STC-BC is impractical for science and engineering problems involving very large data sets [25]. This is due to the poor scaling of inverting the covariance matrix with respect to (w.r.t.) the size of the data sets involved. Assuming we have a data set of size N (including numerical and benchmark data), there will be an N-by-N covariance matrix to invert. The computational complexity to invert an N-by-N matrix is $N^3$. To solve this problem, Higdon et al. [26,27] followed the definition of Kennedy and O'Hagan but suggested to decompose the model output and model discrepancy using singular value decomposition (SVD). The weights generated in the SVD of the numerical output is treated as a function of the model parameters and spatial coordinate. The concept of the STD-BC is then applied to the weight functions. Zhang and Fu [16,17] improved the efficiency of the original Bayesian UQ framework by using the adaptive HDMR technique to decompose the original high dimensional problems into several lower-dimensional subproblems. These treatments subsequentially improves efficiency, however, it does not invoke the online learning capability. Once more benchmark data becomes available, users have to execute, from scratch, the whole procedure with a larger dataset. The work presented in this paper aims to provide a solution to improve the efficiency of the STD-BC and enable the online capability using methods developed within the research literature on Gaussian processes (GP).

GP is a naïve Bayes supervised machine learning algorithm. By assuming the data follows an adjoint multivariate Gaussian distribution, it offers the potential to determine the mean and variances of the unknown. As is the case with the STD-BC, the poor scalability of inversion of the covariance matrix has limited the usage of the GP. In recent years, many efforts had been made to develop a sparse Gaussian processes for very large data sets. A detailed review can be found in [28]. In general, there are several ways to increase the sparsity of a GP, such as the Nyström GP [29,30], deterministic training conditional (DTC) GP [31], and kernel interpolation for scalable structured GP (KISS-GP) [32]. In this paper, we propose to adapt the stochastic variational inference GP (SVI-GP) method [33] to improve the efficiency of the Bayesian calibration and enable the online learning capability. We name the new method 'fixed inducing points online Bayesian calibration' (FIPO-BC).

The rest of this paper is organised as follows: A brief introduction to the SVI-GP and STD-BC is given in section 2 which is followed by the description of FIPO-BC in section 3. In section 4, we summarise the results of a toy case, while section 5 is contributed to calibrate a key model parameter in a scale resolving turbulence model (SAS-SST model). This paper is closed by a summary and discussion of future work in Section 5.

## 2  Background

According to Kennedy and O'Hagan [5], the physical process $f(\cdot)$, noisy observation $y_i \in \mathbb{R}$, and model outputs $h(\cdot,\cdot)$ at $\boldsymbol{x}_i \in \mathbb{R}^d$ has the following form:

$$y_i = f(x_i) + \epsilon_i = h(\boldsymbol{x}_i, \boldsymbol{\vartheta}) + \delta(\boldsymbol{x}_i) + \epsilon_i \qquad (1)$$

where $\epsilon_i$ is the observation error and $\epsilon_i \sim \mathcal{N}(0, \beta_{exp}^{-1})$, $\delta(\cdot)$ is the numerical model inadequacy and $\boldsymbol{\vartheta} \in \mathbb{R}^\theta$ is the vector of model parameters. The Bayesian calibration assumes the numerical dataset and observations is a joint GP and then proceeds to find the most plausible $\boldsymbol{\vartheta}$ via maximise the marginal likelihood.

### 2.1  Stochastic variational inference Gaussian processes (SVI-GP)

Before further description of the Bayesian calibration method, a concise description of Gaussian processes [34,35] and stochastic variational inference Gaussian processes (SVI-GP) [33] are given in this section. This allows us to introduce notation and expressions which will be used in the description of STD-BC and FIPO-BC methods.

The standard GP assumes that any finite subset of $\{f(x)|x \in \mathbb{R}^d\}$ follows an adjoint Gaussian distribution. Considering the vector of noisy observations $y = \{y_i \in \mathbb{R}\}_{i=1}^{n_1}$ taken at $X = \{x_i \in \mathbb{R}^d\}_{i=1}^{n_1}$. In the standard GP, we have

$$p(y|f) = \mathcal{N}(y|f, \beta_{exp}^{-1}I) \tag{1}$$

$$p(f|X) = \mathcal{N}(f|0, K_{ee}) \tag{2}$$

Using Eq.(1 & 2), we define the prior of the real-valued function $f(x)$ as an adjoint multivariant Gaussian distribution with zero means ($0$) and covariance matrix ($K_{ee}$). $K_{ee}$ is an $n_1 \times n_1$ matrix, whose elements are calculated using a predefined covariance function which is typically dependent upon a set of hyperparameters. Hyperparameters can be selected by maximise the marginal likelihood $p(y|X) = \mathcal{N}(y|0, K_{ee} + \beta_{exp}^{-1}I)$. The posterior GP of unknown $y^*$ at $x^*$ is $\mathcal{N}(y^*|K^*K_{ee}^{-1}y, K^{**} - K^*K_{ee}^{-1}(K^*)^T + \beta_{exp}^{-1})$, where $K^*$ is a $1 \times n_1$ covariance vector and $K^{**}$ is calculated at $x^*$ using the covariance function.

The computational complexity of GP is governed by the size of $K_{ee}$. Assuming, we can find inducing variables $u$ that are conditioned on inducing points $Z = \{z_i\}_{i=1}^{m_1}$, preferably with $m_1 \ll n_1$. We then augment the GP prior as

$$p(f|X) = p(f|u, Z, X)p(u|Z) \tag{3}$$

Similar to $p(f|X)$,

$$p(u|Z) = \mathcal{N}(u|0, K_{uu}) \tag{4}$$

, where $K_{uu}$ is a $m_1 \times m_1$ covariance matrix evaluated between all the inducing points. Let $K_{eu}$ (a $n_1 \times m_1$ matrix) be the covariance matrix between $x_{exp}$ and $z$. We will then have,

$$p(f|u, Z, X) = \mathcal{N}(f|K_{eu}K_{uu}^{-1}u, \widetilde{K}) \tag{5}$$

, where $\widetilde{K} = K_{ee} - K_{eu}K_{uu}^{-1}K_{ue}$ and $K_{ue}$ is the transpose of $K_{eu}$. The task then becomes finding the proper $u$ with knowledge of $y$. In the SVI-GP, the variational distribution $q(u) \sim N(\mu, S)$ is introduced to enable the usage of SVI which allows for online training [36]. There is difficulty in determining the optimised inducing inputs ($z$) as it will introduce a dependence on the training data if the optimisation procedure is done w.r.t it. To solve this problem, a fine predifined inducing points can be used [33,37].

Parameters in $q(u)$, including $\mu$, $S$, and hyperparameters of the kernel function, are optimised using the lower boundary of $p(y|X)$. The objection function can be written as

$$L_{SVI-GP} = \sum_{i=1}^{n_1} \left\{ \log N(y_i | \mathbf{k}_i^T \mathbf{K}_{uu}^{-1} \boldsymbol{\mu}, \beta^{-1}) - \frac{\tilde{k}_{ii} + \mathbf{k}_i \mathbf{K}_{uu}^{-1} \mathbf{S} \mathbf{K}_{uu}^{-1} \mathbf{k}_i^T}{2\beta^{-1}} \right\} \quad (6)$$
$$- KL(q(\mathbf{u})|p(\mathbf{u}))$$

with $\mathbf{k}_i$ being the i$^{th}$ row of $\mathbf{K}_{eu}$ and the $\tilde{k}_{ii}$ is the diagonal element of $\widetilde{\mathbf{K}}$. Once $q(\mathbf{u})$ is determined, the posterior GP mean and covariance can be approximated as described in [38].

## 2.2 Standard Bayesian calibration (STD-BC) revisited

Consider observations $\mathbf{y} = \{y_i \in \mathbb{R}\}_{i=1}^{m_1}$ at $\mathbf{X}_{exp} = \{\mathbf{x}_{exp,i} \in \mathbb{R}^d\}_{i=1}^{m_1}$, and simulation outputs $\mathbf{h} = \{h_{i,j} = M(\mathbf{x}_{num,i}, \boldsymbol{\vartheta}_j)\}$ at $\mathbf{X}_{num} = \{\mathbf{x}_{num,i} \in \mathbb{R}^d\}_{i=1}^{n_1}$ and $\boldsymbol{\theta} = \{\boldsymbol{\vartheta}_i \in \mathbb{R}^\theta\}_{i=1}^{a_1}$. The number of elements in $\mathbf{h}$ is $N_1 = a_1 n_1$. It is well known that numerical outputs are also subject to numerical error ($\beta_{num}$). Therefore, we assume $\mathbf{h} \sim \mathcal{N}(\mathbf{0}, \mathbf{L}_{nn})$, in which $\mathbf{L}_{nn} = \mathbf{K}_{num} + \beta_{num}^{-1}\mathbf{I}$ and $\mathbf{K}_{num}$ is a $N_1 \times N_1$ matrix where the element is evaluated using $k_{num}\{(\mathbf{x}_{num,i}, \boldsymbol{\vartheta}_k), (\mathbf{x}_{num,j}, \boldsymbol{\vartheta}_l)\}$. Now, let us assume $\mathbf{y} \sim \mathcal{N}(\mathbf{0}, \mathbf{L}_{ee})$, where $\mathbf{L}_{ee} = \mathbf{K}_{ee} + \beta_{exp}^{-1}\mathbf{I}$ and $\mathbf{K}_{ee} = \mathbf{K}_{num} + \mathbf{K}_\delta$ is the covariance matrix of true values ($\mathbf{f}$). $\mathbf{K}_\delta$ is a $m_1 \times m_1$ matrix for the model inadequacy with elements calculated using $k_\delta(\mathbf{x}_{exp,i}, \mathbf{x}_{exp,j})$. The hyperparameters in the $k_{num}$ and $k_\delta$ are $\boldsymbol{\varphi}_{num}$ and $\boldsymbol{\varphi}_\delta$, respectively. Also, it should be noted that, for simplicity, we imply that the number of grid/mesh in numerical simulations are the same.

Let us denote $\mathbf{d}^T = [\mathbf{h} \quad \mathbf{y}]$. According to the definition of the STD-BC [5], we then have
$$p(\mathbf{d}|\boldsymbol{\vartheta}, \boldsymbol{\varphi}_{num}, \boldsymbol{\varphi}_\delta) = \mathcal{N}(\mathbf{d}|\mathbf{0}, \mathbf{V}_{BB}) \quad (7)$$
In Eq. (7),
$$\mathbf{V}_{BB} = \begin{bmatrix} \mathbf{L}_{nn} & \mathbf{C}_{ne} \\ \mathbf{C}_{en} & \mathbf{L}_{ee} \end{bmatrix} \quad (8)$$
$\mathbf{V}_{BB}$ is $(m_1 + N_1) \times (m_1 + N_1)$ matrix. $\mathbf{C}_{en}$ is a $m_1 \times N_1$ cross-covariance matrix and is designed to describe the relationship between numerical outputs and observations with $\boldsymbol{\vartheta}$. The elements in $\mathbf{C}_{en}$ are evaluated using $k_{num}\{(\mathbf{x}_{i,j}, \boldsymbol{\vartheta}_i), (\mathbf{x}_{exp,k}, \boldsymbol{\vartheta})\}$ and $\mathbf{C}_{ne} = \mathbf{C}_{en}^T$. It should be noted that we eliminated $\mathbf{X} = [\mathbf{X}_{num} \quad \mathbf{X}_{exp}]$ in Eq. (7). It is noted in [5] that marginalised the effect of $\boldsymbol{\varphi}_{num}$ and $\boldsymbol{\varphi}_\delta$ would be too complex. Therefore, the optimal hyperparameters $\boldsymbol{\varphi}_{num}^*$ and $\boldsymbol{\varphi}_\delta^*$ are suggested to be used. It should be noted that $\beta_{num}$ and $\beta_{exp}$ are folded in $\boldsymbol{\varphi}_{num}^*$ and $\boldsymbol{\varphi}_\delta^*$ here.

With fixed hyperparameters, the plausibility of $\boldsymbol{\vartheta}$ is then expresses as

$$p(\boldsymbol{\vartheta}|\boldsymbol{d}, \boldsymbol{\varphi}_{num}^*, \boldsymbol{\varphi}_\delta^*) = \frac{p(\boldsymbol{\vartheta})p(\boldsymbol{d}|\boldsymbol{\vartheta}, \boldsymbol{\varphi}_{num}^*, \boldsymbol{\varphi}_\delta^*)}{\int p(\boldsymbol{\vartheta})p(\boldsymbol{d}|\boldsymbol{\vartheta}, \boldsymbol{\varphi}_{num}^*, \boldsymbol{\varphi}_\delta^*)d\boldsymbol{\vartheta}} \quad (9)$$

Then, the objective in the coupled calibration is to obtain $\underset{\vec{\theta} \in A}{argmax}\, p(\boldsymbol{\vartheta}|\boldsymbol{d}, \boldsymbol{\varphi}_{num}^*, \boldsymbol{\varphi}_\delta^*)$. The denominator in Eq. (9) is a constant and the prior of $\boldsymbol{\vartheta}$ ($p(\boldsymbol{\vartheta})$) should be as flat as possible to avoid strong bias. As a result, the shape and the maxima of $p(\boldsymbol{\vartheta}|\boldsymbol{d}, \boldsymbol{\varphi}_{num}^*, \boldsymbol{\varphi}_\delta^*)$ are effectively determined by $p(\boldsymbol{d}|\boldsymbol{\vartheta}, \boldsymbol{\varphi}_{num}^*, \boldsymbol{\varphi}_\delta^*)$. Then $\underset{\vec{\theta} \in A}{argmax}\, p(\boldsymbol{\vartheta}|\boldsymbol{d}, \boldsymbol{\varphi}_{num}^*, \boldsymbol{\varphi}_\delta^*)$ becomes $\underset{\boldsymbol{\vartheta} \in A}{argmax}\, p(\boldsymbol{d}|\boldsymbol{\vartheta}, \boldsymbol{\varphi}_{num}^*, \boldsymbol{\varphi}_\delta^*)$. The logarithm form of $p(\boldsymbol{d}|\boldsymbol{\vartheta}, \boldsymbol{\varphi}_{num}^*, \boldsymbol{\varphi}_\delta^*)$ can be written as

$$log\, p(\boldsymbol{d}|\boldsymbol{\vartheta}, \boldsymbol{\varphi}_{num}^*, \boldsymbol{\varphi}_\delta^*) = -\frac{1}{2}\boldsymbol{d}^T \boldsymbol{V}_{BB}^{-1} \boldsymbol{d} - \frac{1}{2} log|\boldsymbol{V}_{BB}| - \frac{m_1 + N_1}{2} log 2\pi \quad (10)$$

The STD-BC is then to solve $\underset{\boldsymbol{\vartheta} \in A}{argmax}\, log\big(p(\boldsymbol{d}|\boldsymbol{\vartheta}, \boldsymbol{\varphi}_{num}^*, \boldsymbol{\varphi}_\delta^*)\big)$ in which $A \subseteq \mathbb{R}^\theta$. For the sake of simplicity, we will exclude $\boldsymbol{\varphi}_{num}^*$ and $\boldsymbol{\varphi}_\delta^*$ in Eq.(10) and write $log\big(p(\boldsymbol{d}|\boldsymbol{\vartheta}, \boldsymbol{\varphi}_{num}^*, \boldsymbol{\varphi}_\delta^*)\big)$ as $log\big(p(\boldsymbol{d}|\boldsymbol{\vartheta})\big)$ in the following discussion.

The flow chart for STD-BC is illustrated in the Figure 1. STD-BC contains three major steps:

1. Determine a set of optimal hyperparameters ($\boldsymbol{\varphi}_{num}^*$) for $k_{num}$ w.r.t. the marginal likelihood of $\boldsymbol{h} \sim \mathcal{N}(\boldsymbol{0}, \boldsymbol{L}_{nn})$.
2. Determine a set of optimal hyperparameters ($\boldsymbol{\varphi}_\delta$) for $k_\delta$ with knowledge of $\boldsymbol{\varphi}_{num}^*$.
3. Solve $\underset{\boldsymbol{\vartheta} \in A}{argmax}\, ln\big(p(\boldsymbol{d}|\boldsymbol{\vartheta})\big)$ to obtain calibrated $\boldsymbol{\vartheta}$ or to obtain the posterior distribution of $\boldsymbol{\vartheta}$ w.r.t. $p(\boldsymbol{d}|\boldsymbol{\vartheta})$ for the inverse uncertainty quantification purpose.

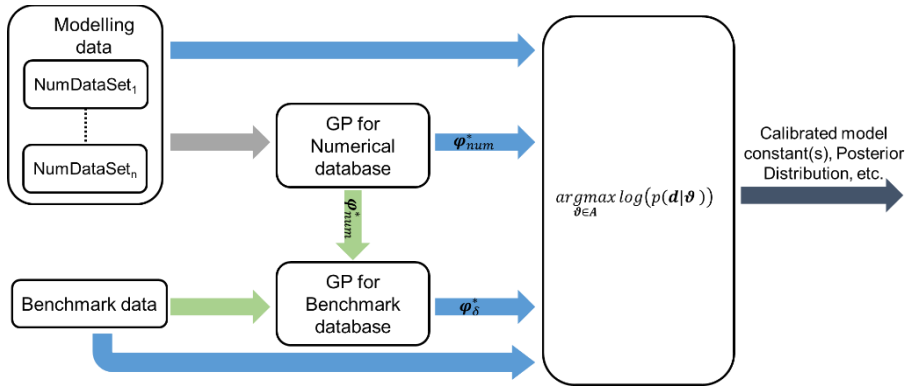

Figure 1 Flow chart of the standard Bayesian calibration (STD-BC).

## 3 Fixed inducing points online Bayesian calibration (FIPO-BC)

In the FIPO-BC framework, an extra layer formed by the inducing points is introduced to eliminate the dependence of the size of the covariance matrix on available data. The form of the likelihood defined in the Bayesian calibration is applied to the variables on the inducing points. The likelihood is then marginalised w.r.t the variational distribution of the inducing variables which is obtained using the SVI method.

Let $\tilde{\boldsymbol{u}} = \{u_i = f(\boldsymbol{z}_{exp,i}), \boldsymbol{z}_{exp,i} \in \mathbb{R}^d\}$ and $\boldsymbol{W} = \{w_{i,j} = M(\boldsymbol{z}_{num,i}, \boldsymbol{\omega}_j), \boldsymbol{z}_{num,i} \in \mathbb{R}^d \text{ and } \boldsymbol{\omega}_j \in \mathbb{R}^\theta\}$ to be the inducing variables for $\boldsymbol{f}$ and $\boldsymbol{h}$. Let $\boldsymbol{Z}_{exp} = \{\boldsymbol{z}_{exp,i} \in \mathbb{R}^d\}_{i=1}^{m_2}$, $\boldsymbol{Z}_{num} = \{\boldsymbol{z}_{num,i} \in \mathbb{R}^d\}_{i=1}^{n_2}$, $\boldsymbol{\Omega} = \{\boldsymbol{\omega}_j \in \mathbb{R}^\theta\}_{j=1}^{a_2}$, and $\boldsymbol{Z} = [\boldsymbol{Z}_{num} \quad \boldsymbol{Z}_{exp}]$. So $\boldsymbol{W}$ contains $N_2 = a_2 n_2$ elements and $N_2 + m_2 \ll N_1 + m_1$. Let $\tilde{\boldsymbol{w}}$ be the vectorised $\boldsymbol{W}$ and $\boldsymbol{d}_S^T = [\tilde{\boldsymbol{w}} \quad \tilde{\boldsymbol{u}}]$. Similar to the STD-BC, we now define the $p(\boldsymbol{d}_S|\boldsymbol{\vartheta})$ as

$$p(\boldsymbol{d}_S|\boldsymbol{\vartheta}, \tilde{\boldsymbol{\varphi}}_{num}^*, \tilde{\boldsymbol{\varphi}}_\delta^*) = \mathcal{N}(\boldsymbol{d}_S|\boldsymbol{0}, \boldsymbol{V}_{SS}) \tag{11}$$

, and

$$\boldsymbol{V}_{SS} = \begin{bmatrix} \boldsymbol{L}_{ww} & \boldsymbol{C}_{wu} \\ \boldsymbol{C}_{uw} & \boldsymbol{L}_{uu} \end{bmatrix} \tag{12}$$

In Eq. (12), $\boldsymbol{L}_{ww} = \boldsymbol{K}_{ww}$ and $\boldsymbol{L}_{uu} = \boldsymbol{K}_{uu}$. $\boldsymbol{K}_{ww}$ and $\boldsymbol{K}_{uu} = \boldsymbol{K}_{ww} + \boldsymbol{K}_{\delta'}$ are the covariance matrix for $\tilde{\boldsymbol{w}}$ and $\tilde{\boldsymbol{u}}$, respectively. $\boldsymbol{K}_{\delta'}$ is the covariance matrix with element calculated using kernel function $k_\delta$ for the model inadequacy related to $\tilde{\boldsymbol{u}}$. $\boldsymbol{C}_{uw}$ is the cross-covariance matrix for $\tilde{\boldsymbol{w}}$ and $\tilde{\boldsymbol{u}}$ with element evaluated using $k_{num}$ and $\boldsymbol{C}_{wu} = \boldsymbol{C}_{uw}^T$.

We further write the variational distribution of $\tilde{\boldsymbol{w}}$ and $\tilde{\boldsymbol{u}}$ to be

$$q(\tilde{\boldsymbol{w}}) = \mathcal{N}(\tilde{\boldsymbol{w}}|\boldsymbol{M}_{\tilde{\boldsymbol{w}}}, \boldsymbol{S}_{ww}) \tag{13}$$

$$q(\tilde{\boldsymbol{u}}) = \mathcal{N}(\tilde{\boldsymbol{u}}|\boldsymbol{M}_{\tilde{\boldsymbol{u}}}, \boldsymbol{S}_{uu}) \tag{14}$$

Let $\boldsymbol{M}_S^T = [\boldsymbol{M}_{\tilde{\boldsymbol{w}}} \quad \boldsymbol{M}_{\tilde{\boldsymbol{u}}}]$ and $\boldsymbol{S} = \begin{bmatrix} \boldsymbol{S}_{ww} & \boldsymbol{0} \\ \boldsymbol{0} & \boldsymbol{S}_{uu} \end{bmatrix}$, then

$$q(\boldsymbol{d}_S) = \mathcal{N}(\boldsymbol{d}_S|\boldsymbol{M}_S, \boldsymbol{S}) \tag{15}$$

Expectation of $p(\boldsymbol{d}_S|\boldsymbol{\vartheta}, \tilde{\boldsymbol{\varphi}}_{num}^*, \tilde{\boldsymbol{\varphi}}_\delta^*)$ given $q(\boldsymbol{d}_S)$ is

$$\langle p(\boldsymbol{d}_S|\boldsymbol{\vartheta}, \widetilde{\boldsymbol{\varphi}}^*_{num}, \widetilde{\boldsymbol{\varphi}}^*_\delta)\rangle_{q(\boldsymbol{d}_S)}$$

$$= \frac{1}{\sqrt{(2\pi)^{m_2+N_2}|\boldsymbol{V}_{SS}+\boldsymbol{S}|}} \exp\left(-\frac{1}{2}\boldsymbol{M}_s^T(\boldsymbol{V}_{SS}+\boldsymbol{S})^{-1}\boldsymbol{M}_s\right) \quad (16)$$

$$= \mathcal{N}(\boldsymbol{M}_s|\boldsymbol{0}, \boldsymbol{V}_{SS}+\boldsymbol{S})$$

By eliminating the hyperparameters in the expression, we can now comfortably write $\langle p(\boldsymbol{d}_S|\boldsymbol{\vartheta}, \widetilde{\boldsymbol{\varphi}}^*_{num}, \widetilde{\boldsymbol{\varphi}}^*_\delta)\rangle_{q(\boldsymbol{d}_S)}$ as $p(\boldsymbol{M}_S|\boldsymbol{\vartheta})$ and its logarithm form is

$$\log(p(\boldsymbol{M}_S|\boldsymbol{\vartheta}))$$

$$= -\frac{1}{2}\boldsymbol{M}_s^T(\boldsymbol{V}_{SS}+\boldsymbol{S})^{-1}_{BB}\boldsymbol{M}_s \quad (17)$$

$$-\frac{1}{2}\log|\boldsymbol{V}_{SS}+\boldsymbol{S}| - \frac{m_2+N_2}{2}\log(2\pi)$$

Hyperparameters ($\widetilde{\boldsymbol{\varphi}}^*_{num}$ and $\widetilde{\boldsymbol{\varphi}}^*_\delta$) and $q(\boldsymbol{d}_S)$ can be continuously updated according to the new training data, therefore enables online learning. The flow chart of the FIPO-BC is illustrated in Figure 2. The proposed FIPO-BC algorithm contains following steps:

1. Design the inducing points;
2. Determine a robust set of hyperparameters ($\widetilde{\boldsymbol{\varphi}}^*_{num}$) and infer the model output ($\boldsymbol{M}_{\widetilde{w}}$) at the inducing points using the SVI-GP regarding to the whole set/a batch of model outputs;
3. Determine a robust set of hyperparameters ($\widetilde{\boldsymbol{\varphi}}^*_\delta$) and infer the measurements ($\boldsymbol{M}_{\widetilde{u}}$) at the inducing points using the SVI-GP regarding to the whole set/a batch of the benchmark database;
4. Obtain the calibrated model parameters by solving $\underset{\boldsymbol{\vartheta}\in A}{argmax}\ log(p(\boldsymbol{M}_S|\boldsymbol{\vartheta}))$ or obtain the posterior probability distribution of $\boldsymbol{\vartheta}$ w.r.t $p(\boldsymbol{M}_S|\boldsymbol{\vartheta})$.
5. Update the calibrated $\boldsymbol{\vartheta}$ and posterior probability distribution of $\boldsymbol{\vartheta}$ with knowledge ($\boldsymbol{M}_{\widetilde{w}}$, $\boldsymbol{S}_{ww}$, $\boldsymbol{M}_{\widetilde{u}}$, and $\boldsymbol{S}_{uu}$) obtained in the previous FIPO-BC process, once more data is available.

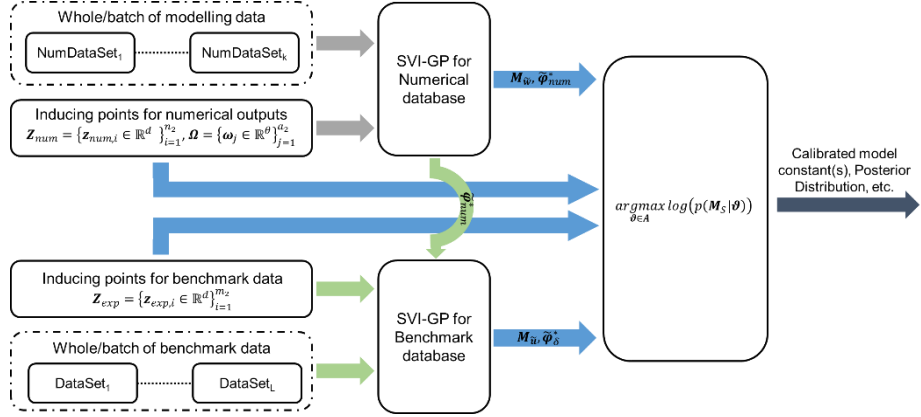

Figure 2 Flow chart of the fixed inducing points online Bayesian calibration framework.

## 4  Application of the FIPO-BC algorithm to a simple test case

This section verifies the implementation as well as demonstrating the practical application of the FIPO-BC algorithm by determining the parameter within a simple mathematical function. Both the STD-BC and FIPO-BC algorithms are used to identify the parameter in the mathematical function $f(x) = x\sin(4.0x)$. To distinguish the true function and model, the latter is expressed as $h(x, C) = x\sin(Cx)$. Moreover, forty measurements are manufactured by adding noise ($\epsilon \sim N(0, 0.05^2)$) to the function outputs. Outputs of $h(x, C)$ are obtained using different values for $C$ but the same mesh. Twenty constants are sampled in the uniform distribution $U(0.0, 8.0)$ using the Latin hypercube sampling (LHS) method. Manufactured measurements and model outputs together with the profile of $f(x)$ are presented in *Figure 3*(a). Extra validation data is generated to validate the hyperparameters of GP models in the calibration process. The validation dataset, shown in *Figure 3*(b), contains twenty measurements and outputs of $h(x, C)$ with ten $C$ values sampled in the uniform distribution $U(0.0, 8.0)$.

Four FIPO-BC cases with different sets of inducing point are generated. The details of the sets of the inducing points are documented in Table 1. From Case 1 to Case 4, the number of the inducing points gradually increases. Outputs of FIPO-BC algorithm, including the calibrated model parameter, the posterior distribution of the parameter, and the profile of the objective function, are compared to those of STD-BC algorithm. The hyperparameters of the surrogate models are optimised using the adaptive Nelder-Mead algorithm [39]. The validity of the GP/SVI-GP models for $h(x, C)$ and measurements ($y$) are checked using the validation dataset. For the sake of simplicity, the GP model is expressed as $g$ and the SVI-GP model is expressed as $g_s$. Several attempts to determine the optimal hyperparameters have been done. In each attempt, the initial searching point of the adaptive Nelder-Mead simplex algorithm is carefully selected. Only the best-performed model is considered in the following discussion.

The quantity-to-quantity (Q-Q) plot and the mean square error (MSE) are illustrated in Figure 4. Outputs of $g_s(x, C)$ are very similar to those of $h(x, C)$ except for Case 1, seeing Figure 4(a). $g_s(x, C)$ in Case 1 is visibly different from the validation data with MSE $\cong$ 1.0E-1. The MSEs of $g_s(x, C)$ in Cases 2, 3 & 4 are smaller than 1.0E-4, while the MSE of $g(x, C)$ in standard BC (STD-BC) is just above 1.0E-4. Furthermore, $g_s(x)$ and $g(x)$ for the manufactured measurement almost overlap with each other, referring to Figure 4(b). The MSEs (<1.0E-02) of $g_s(x)$ and $g(x)$ are much smaller than the standard deviation (0.05) of $\epsilon$. It is confident to claim that the surrogate models in the STD-BC case and last three FIPO-BC cases are valid.

The posterior distribution of $C$ is obtained using the Metropolis-Hasting (MH) algorithm, by assuming that the prior distribution of $C$ follows a uniform distribution in [0, 8]. 10,000 samples have been taken in each case. The histogram of the posterior distribution of $C$, the profile of log marginal likelihood defined in both FIPO-BC and STD-BC, as well as maxima of the profiles are plotted in Figure 5. Because of the deficiency of $g_s(x, C)$ in Case 1, the calibrated $C$ is well away from 4.0. In other FIPO-BC cases and STD-BC, calibrated $C$ is the same as the true value. With more inducing points added in the FIPO-BC, the profile of $log(p(M_S|\vartheta))$ becomes more and more akin to the profile of $log(p(d|\vartheta))$ until the FIPO-BC is converged on the inducing points, i.e. outputs of Case 4 is similar to outputs of Case 3. However, the posterior distribution of $C$ in Case 4 is still visibly different from the outcome of STD-BC. This difference may be due to the non-marginalised hyperparameter effect.

The CPU time per 1000 MH sampling in each case is presented in Figure 6. The in-house Matlab code runs on a laptop with the Intel i7-7600U CPU. The time efficiency of MH sampling is greatly improved when using FIPO-BC algorithm. For instance, the CPU time per 1000 MH sampling in Case 3 is 135.8 s just a tenth of that in the STD-BC algorithm.

In order to demonstrate the online capability of the proposed method, the output of $h(x, C)$ in the training dataset is split into four batches. Each batch is separately fed into the FIPO-BC solver of Case 3. The results are shown in Figure 7, which includes the value of $C$ used to generate the data in the batches, response surface of true model output approximated using the $g_s(x, C)$, as well as the calibrated value of $C$ at each step. With more data fed in the SVI-GP solver, the response surface gradually converges, and therefore to the calibrated value of $C$. The calibrated $C$ with the first set of data ('Batch-1') is 3.92. With the 'Batch-2' fed into the calibration, while 'Batch-1' is omitted, the calibrated $C$ is 4.0. Although the calibrated $C$ shifted to 4.01 with 'Bath-3' and 'Bath-4', such a small difference does not cause much difference in the function output. Therefore, it is reasonable to claim that the online FIPO-BC converges with dataset 'Batch-2'.

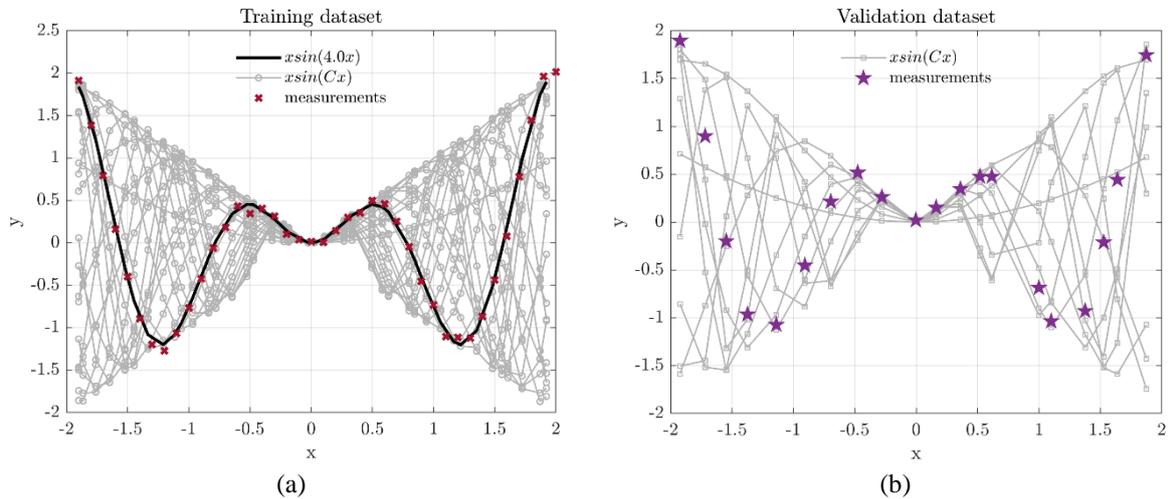

(a)                                                    (b)

Figure 3 (a) Training dataset: (model outputs with different constant ($xsin(Cx)$) and manufactured measurements (benchmark database)); (b) Verification dataset.

Table 1 number of points in the standard Bayesian calibration and resolution of fixed inducing points online Bayesian calibration.

| Cases: | C | x | Measurements | Data points/Inducing points |
|---|---|---|---|---|
| STD-BC: | | | | |
| Standard BC | 20 | 40 | 40 | 840 |
| FIPO-BC: | | | | |
| Case 1 | 6 | 11 | 16 | 82 |
| Case 2 | 9 | 16 | 21 | 165 |
| Case 3 | 11 | 21 | 26 | 257 |
| Case 4 | 17 | 26 | 31 | 473 |

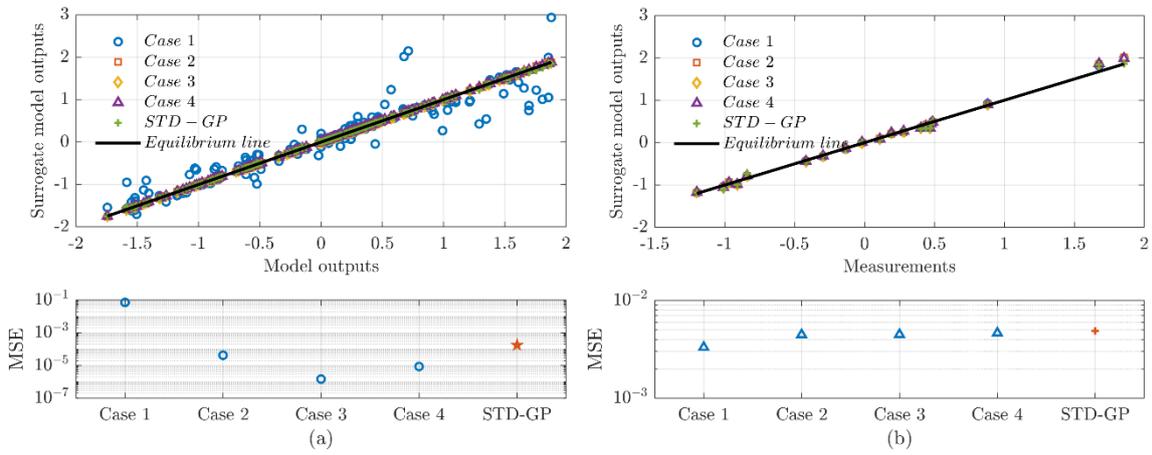

Figure 4 Validation for (a) surrogate models for $h(x, C)$ (b) surrogate models for measurements in FIPO-BC and STD-BC.

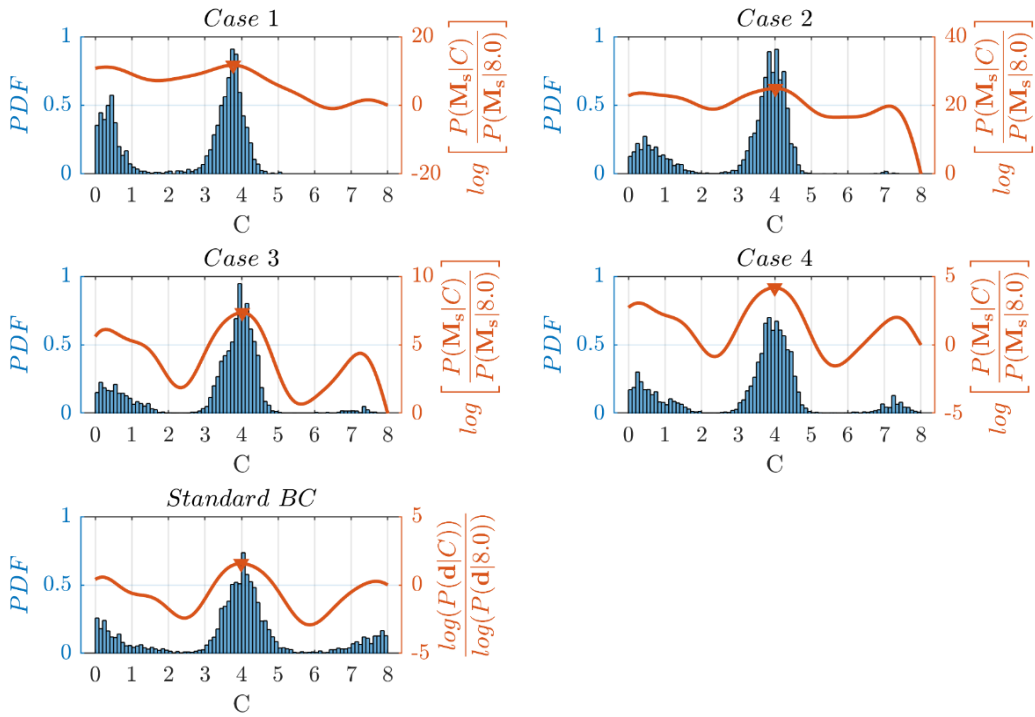

Figure 5 The posterior PDF of $C$ and the log marginal likelihood for cases of FIPO-BC (a) Case 1; (b) Case 2; (c) Case 3; (d) Case 4 and (e) the STD-BC case.

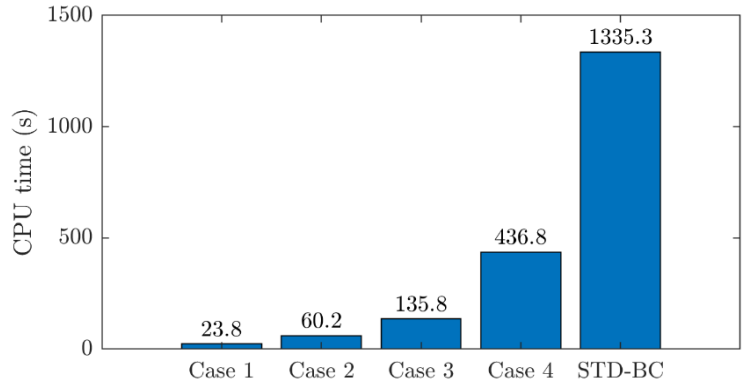

Figure 6 CPU time(s) per 1000 MH sampling in each case.

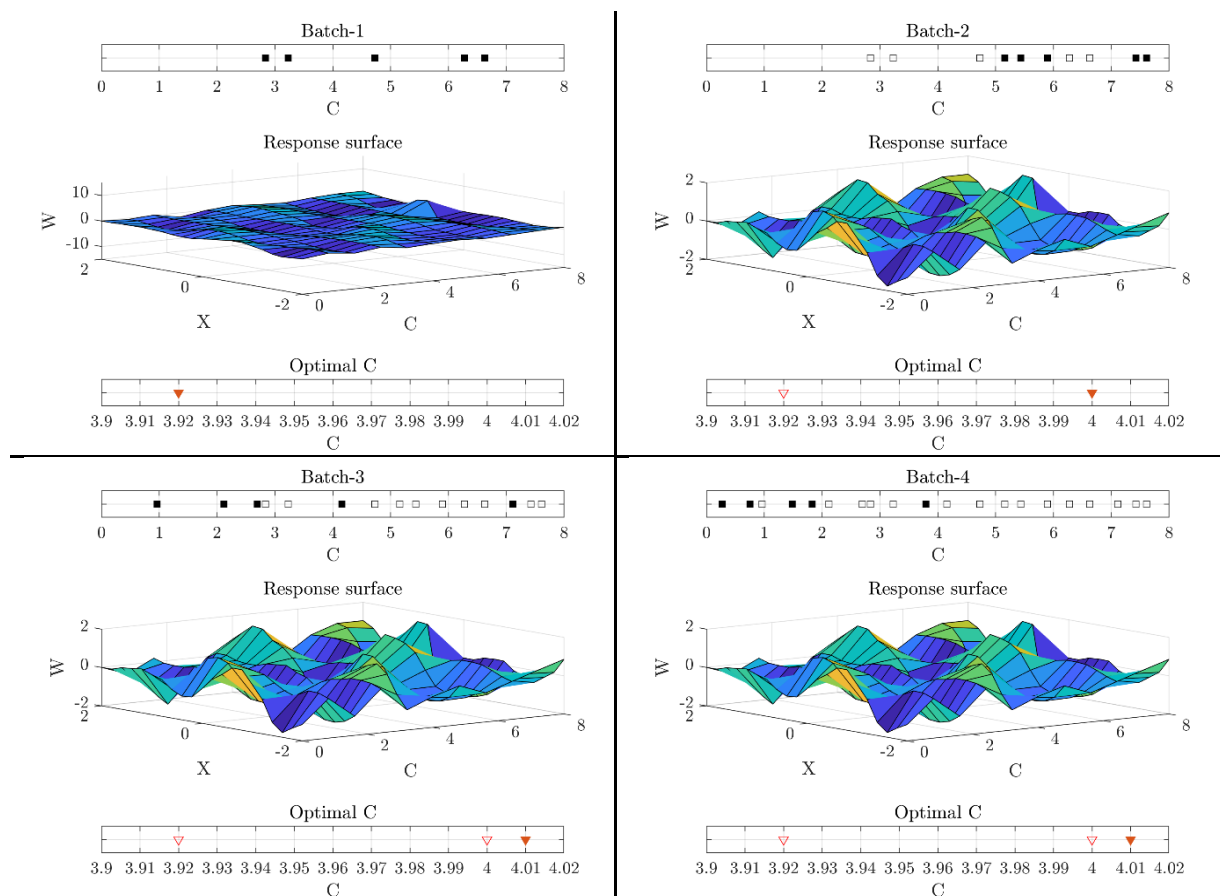

Figure 7 The online calibration of the parameter in the toy case using FIPO-BC. Each pane shows the response surface of the inducing output for the models as well as the calibrated parameter after a batch of model output using different $C$ values, marked as solid points. Previously used (and discarded) $C$ values and the calibrated constant are marked as empty points.

## 5 Calibrating $C_s$ in a SAS-SST simulation

Turbulence modelling has been a continuous endeavour since the 1950s. The challenge is caused by the misalignment of the scale of a flow system and the turbulent eddies with the spatial scale of micrometre/millimetre and the time scale of microsecond/millisecond. Despite their small scale, the turbulent eddies dominate the behaviour of the turbulent flow. Major efforts have been spent in the development of turbulence models which describe the effect of turbulent eddies sufficiently well. In this section, the FIPO-BC algorithm is used to calibrate a damping factor ($C_s$) in a scale-resolving CFD turbulence model (SAS-SST). The SAS-SST model [40,41] aims to generate turbulent eddies using information obtained from the mean flow. Unrealistic turbulent eddies in the high-wave number region may be introduced into the simulated turbulent flow system. As a result, a specific term is designed to filter those unphysical turbulent eddies. The damping factor ($C_s$) controls the size of the filter and is the key parameter in this term. It changes the balance of the modelled and resolved turbulence effect, and hence affecting the model accuracy.

The FIPO-BC algorithm is applied to calibrate the $C_s$ with regard to the streamwise velocity measurement of incompressible flow passing a prism bluff body [42,43]. The SAS-SST model is implemented within the commercial CFD solver, STAR-CCM+ 12.04, using user-defined functions. Eleven SAS-SST simulations are performed to provide the database for the surrogate model of the numerical simulation. Except for $C_s$, other settings, such as boundary conditions and numerical set-up, are the same in these simulations. Further details about how the simulations are set-up can be found in [4]. The chosen $C_s$ are 0.0, 0.1, …, 1.0. Each simulation takes more than 96,000 CPU hours to reach time-convergence. Experimental measurements and SAS-SST predictions of $U/U_b$ just after the bluff-body are plotted in Figure 8(a). After comparing the SAS-SST outputs to measurements graphically, it is clear that the SAS-SST with $C_s = 0.8, \ 0.9 \ and \ 1.0$ better matches the measurements. Hereby, we plot the mean square difference (MSD) between SAS-SST prediction using $C_s = 0.9$ and other SAS-SST predictions in Figure 8 (b). The MSD of simulation using $C_s = 1.0$ is 6.6E-6. It suggests the SAS-SST using $C_s$ between 0.9 and 1.0 would produce almost the same $U/U_b$ at the location.

The FIPO-BC with several different sets of inducing points are performed of which details are included in Table 2. In the first three cases, the number of the inducing points for $C_s$ gradually increases from 6 to 11, while it is kept the same (36) on the spatial coordinate ($X$).

Spatial inducing points increase to 41 and 46 in the last two cases, respectively. Again, the adaptive NM simplex algorithm is chosen to obtain the hyperparameters. But in this part of the study, the initial searching points are randomised. Dozens of attempts have been made and the hyperparameters resulting in the best results are chosen to show here.

Suitability checking of the surrogate models for CFD models and measurements are illustrated in Figure 9. As in the toy case, the surrogate models in the FIPO-BCs are checked by using the inducing data points to predicted the training dataset, whilst the LOO-CV test is adopted to assess the validity of surrogate models in the STD-BC. As shown in Figure 9 (a), a visible improvement of surrogate model accuracy is achieved by refining the inducing points. Compared to the other FIPO-BC cases, data points produced in FIPO-BC-5 are closer to the equilibrium line in the quantity-to-quantity plot. With a small number of inducing points for spatial coordinate, the improvement on SVI prediction of CFD outputs due to refined $C_s$ resolution is too small to be seen, referring to the MSE of FIPO-BC-1, 2, 3. As the number of inducing points for spatial coordinate increases, the MSE drops from ~4.0E-4 to ~9.0E-5 in FIPO-BC-4 and 5. Additionally, the surrogate model for CFD models in the STD-BC also performs well (MSE≅1.0E-4). As shown in Figure 9 (b), the SVI-GP model for measurement converges at FIPO-BC-2, containing 16 inducing points. The MSEs of SVI-GP predictions of measurements are around 3.0E-3 in FIPO-BC cases, except for FIPO-BC-1. Interestingly, the STD-GP model for the measurements in the STD-BC shows worse accuracy comparing to the SVI grid converged SVI-GP models, as its MSE is ~2.0E-02. This may be due to the sparsely distributed noisy data points and the LOO-CV test.

Figure 10 illustrates the histogram of the posterior PDF of $C$ (obtained using the MH algorithm), as well as profiles of log marginal likelihood and maximums in FIPO-BC cases and STD-BC are plotted. As the inducing points refined, the shape of $log(p(\boldsymbol{M}_s|C_s))$ becomes more like that of $log(p(\boldsymbol{d}|C_s))$. Since the effect of hyperparameters is not marginalised, the range of the marginal likelihood greatly varies and causes differences in the posterior PDF. For instance, the range of $log(p(\boldsymbol{M}_s|C_s))$ in FIPO-BC-2 and 4 are much larger than that in the other cases. Subsequently, the posterior distribution of $C_s$ in these three cases is more concentrated than others. The maximum values in the log marginal likelihood of all cases are 1.0, 0.83, 0.89, 0.92, 0.92, and 0.9 for FIPO-BC-1, 2, 3, 4, 5 and STD-BC respectively. Since the SAS-SST prediction using $C_S \in [0.9, 1.0]$ can be deemed to be similar due to the very small MSD between them. It is reasonable to claim the inducing points convergence is achieved at FIPO-BC-4. Figure 11 illustrates the computing efficiency (represented by CPU time(s) per 1000 MH sampling) of different cases. As expected, the time spent on MH sampling greatly reduces in all FIPO-BC cases. Even for FIPO-BC-5 (with finest SVI grid), CPU time(s) per 1000 MH sampling is less than a tenth of that of the STD-BC.

To demonstrate the online FIPO-BC, the original data is split into four different data batches and fed into the FIPO-BC solver sequentially, whilst the inducing is the same as that used in FIPO-BC-5. The results of this online calibration are shown in Figure 12. Each pane of Figure 12 includes the $C_s$ used to generate the numerical outputs, response surface of absolute error (|numerical output – SVGP prediction|), as well as the optimal $C_s$. Clearly, with more data fed into the FIPO-BC solver, the absolute error is reduced across the parameter space. Meanwhile, the optimal $C_s$ continuously improves. The optimal $C_s$ obtained using 'Batch-1, 2, 3 & 4' are 0.5, 1.0, 1.0, and 0.94, respectively. In particular, the absolute errors produced at the stage of 'Batch-3' and 'Batch-4' are very similar. We can confidently claim that the calibration can be achieved only using Batch-1, 2 & 3 when the FIPO-BC is used.

Table 2 Numbers of the inducing points for calibrating $C_s$ in SAS-SST using FIPO-BC and numbers of data point in the STD-BC.

| Cases: | $C_s$ | x-coordinate | Measurements | Inducing Points / Data points |
|---|---|---|---|---|
| FIPO-BC-1 | 6 | 36 | 11 | 227 |
| FIPO-BC-2 | 9 | 36 | 16 | 340 |
| FIPO-BC-3 | 11 | 36 | 21 | 417 |
| FIPO-BC-4 | 11 | 41 | 26 | 477 |
| FIPO-BC-5 | 11 | 46 | 26 | 532 |
| STD-BC | 11 | 121 | 54 | 1385 |

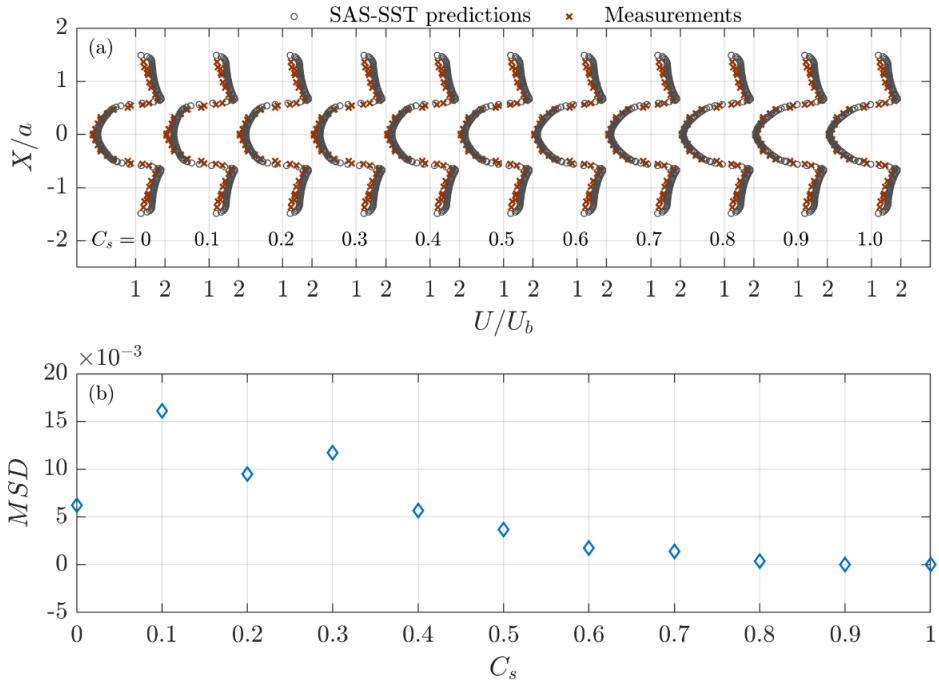

Figure 8 (a) Profiles of $U/U_b$ after the prism bluff-body predicted using SAS-SST model with different $C_s$, (b) mean square difference of $U/U_b$ between different SAS-SST simulation and results obtained using $C_s = 0.9$.

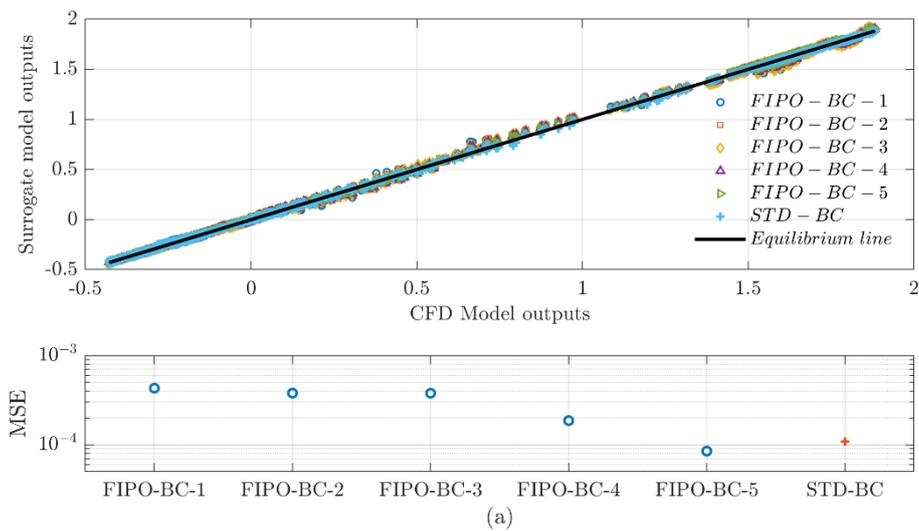

(a)

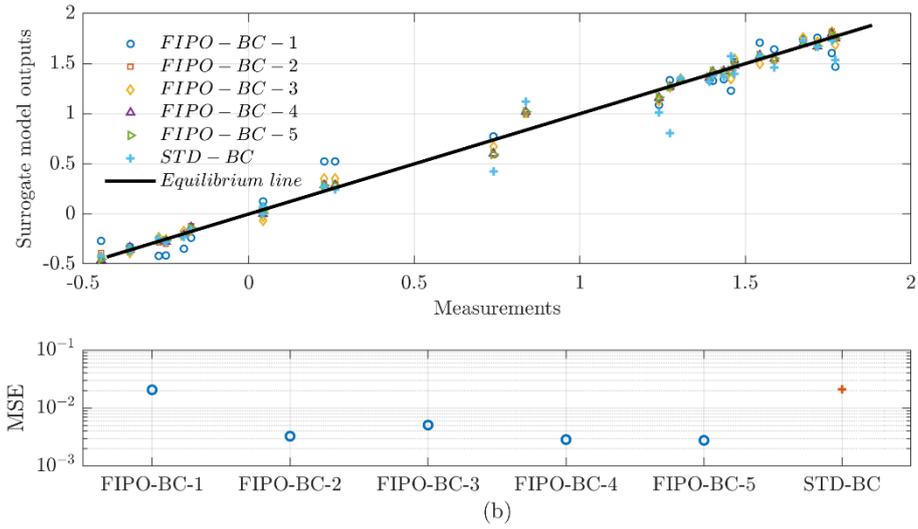

Figure 9 Validation of surrogate models for (a) CFD outputs (b) measurements in FIPO-BC cases and STD-BC (Leave-one-out test is adopted to validate the surrogate models in STD-BC).

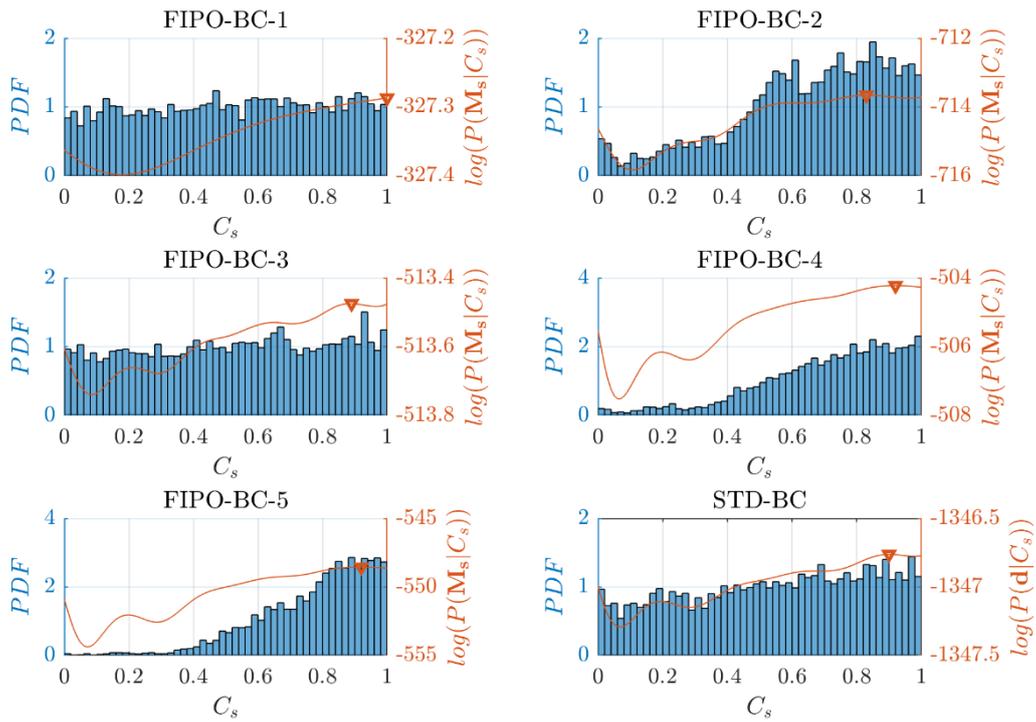

Figure 10 The posterior PDF of $C_s$ and the distributions log marginal likelihood for the FIPO-BC cases and the STD-BC case.

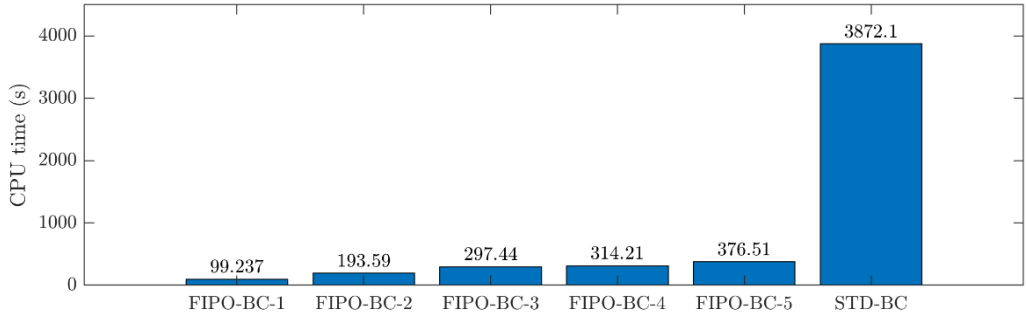

Figure 11 CPU time(s) per 1000 MH sampling in each case for exploring posterior distribution of $C_s$.

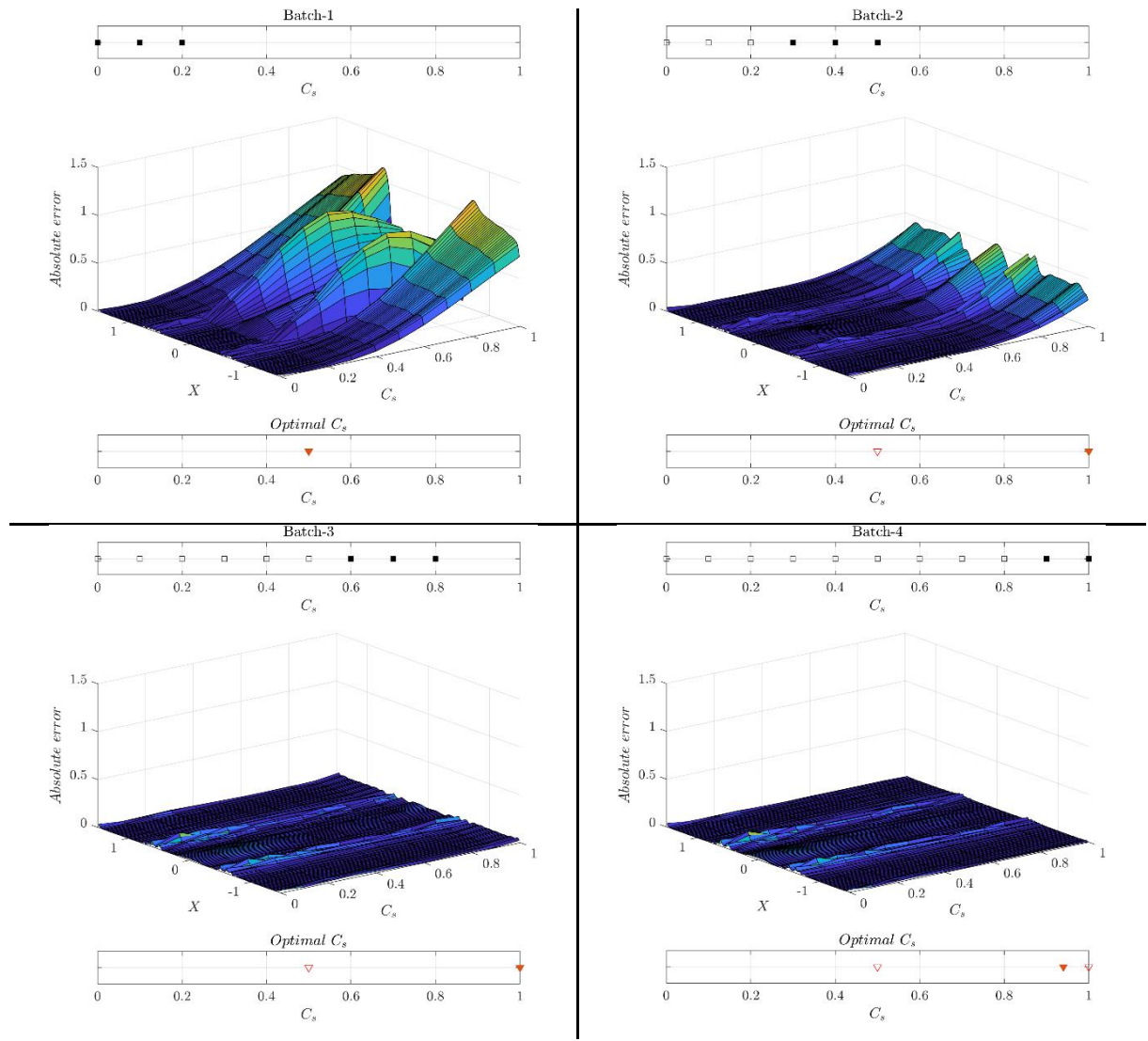

Figure 12 The online calibration of $C_s$ in SAS-SST model. Each pane shows the response surface of the inducing output for the models as well as the calibrated $C_s$ after a batch of model output using different $C_s$, marked as solid points. The $C_s$ used to generate model

outputs considered in the previous step (but discarded the current step) and previously obtained calibrated value are marked as empty points.

## 6  Summary and Discussion

We have proposed a novel fixed inducing points online Bayesian calibration framework (FIPO-BC) algorithm for improving the accuracy of numerical simulations of science and engineering problems. The framework improves the computational efficiency and enables the online learning capability of the STD-BC algorithm by incorporating the SVI-GP. The FIPO-BC algorithm can use far fewer data points, resulting in a much smaller covariance matrix, and hence producing a much more computationally efficient algorithm. The new approach also allows online updates of the calibrated model parameter with data being supplied to the solver sequentially. The procedure of using the FIPO-BC algorithm is demonstrated by using it first to calibrate the parameter in a simple mathematical function, and then to calibrate an essential parameter of scale-resolving CFD modelling method, SAS-SST. The performance of the FIPO-BC algorithm is compared against the STD-BC algorithm.

It has been demonstrated in both test cases that FIPO-BC algorithm provides the same calibrated parameter as the STD-BC algorithm, once the FIPO-BC algorithm is converged on the number of inducing points. Due to the online learning capability of the FIPO-BC algorithm, the calibration can be achieved using less data. Like the STD-BC algorithm, the posterior distribution of the parameter will be affected by the optimal hyperparameters. Both the STD-BC and FIPO-BC algorithms require marginalisation to eliminate this effect. In order to enable the algorithms to utilise marginalisation requires additional computational resources. However, the inclusion of marginalisation within the FIPO-BC algorithm would be more computationally efficient than the STD-BC algorithm with marginalisation. The FIPO-BC algorithm with marginalisation will be investigated in future work.

## Acknowledgements

The authors would like to thank Rolls-Royce for the financial support via the project grant P65165_MECM.